\documentclass[conference]{IEEEtran}
\IEEEoverridecommandlockouts
\def\BibTeX{{\rm B\kern-.05em{\sc i\kern-.025em b}\kern-.08em
    T\kern-.1667em\lower.7ex\hbox{E}\kern-.125emX}}

\usepackage{caption}
\usepackage{amsmath,graphicx}
\usepackage{float}

\usepackage{etoolbox}

\usepackage{amssymb}
\usepackage{amsmath}
\usepackage{amsthm}
\usepackage{booktabs}
\usepackage{enumitem}
\usepackage{graphicx}
\usepackage{color}
\usepackage{algpseudocode}
\usepackage{times}
\usepackage{epsfig}
\usepackage{graphicx}
\usepackage{amssymb}
\usepackage{booktabs}
\usepackage{multirow}
\usepackage{verbatim}
\usepackage{cite}
\usepackage{hyperref}
\usepackage{lscape}
\usepackage{tabularx}
\usepackage{amssymb}
\usepackage{hyperref}
\usepackage{gensymb}
\usepackage{subfig}
\usepackage{color}
\usepackage{colortbl}
\usepackage{pifont}
\usepackage{microtype}

\usepackage[dvipsnames,svgnames,x11names,table]{xcolor}
\usepackage{caption}
\usepackage{enumitem}
\usepackage{booktabs}
\usepackage{algorithm}
\usepackage{pdfpages}
\usepackage{algpseudocode}
\usepackage{adjustbox}
\usepackage{amsmath} 
\usepackage{siunitx}
\usepackage{booktabs}
\usepackage{multirow}
\floatname{algorithm}{Algorithm} 

\newcommand{\dorowcolors}{\rowcolors{2}{gray!15}{white}}

\title{SHAP-Integrated Convolutional Diagnostic Networks for Feature-Selective Medical Analysis}

\author{ \IEEEauthorblockN{Yan Hu\IEEEauthorrefmark{1}, Ahmad Chaddad\IEEEauthorrefmark{1}\IEEEauthorrefmark{2}}
  
  \IEEEauthorblockA{\IEEEauthorrefmark{1}Laboratory for AIPM, School of Artificial Intelligence, Guilin University of Electronic Technology, China}
  \IEEEauthorblockA{\IEEEauthorrefmark{2}The Laboratory for Imagery, Vision and Artificial Intelligence, École de Technologie Supérieure, Canada\\Email: ahmad8chaddad@gmail.com \\ }
}

\begin{document}
%
\maketitle
\begin{abstract}

This study introduces the SHAP-integrated convolutional diagnostic network (SICDN), an interpretable feature selection method designed for limited datasets, to address the challenge posed by data privacy regulations that restrict access to medical datasets. The SICDN model was tested on classification tasks using pneumonia and breast cancer datasets, demonstrating over 97\% accuracy and surpassing four popular CNN models. We also integrated a historical weighted moving average technique to enhance feature selection. The SICDN shows potential in medical image prediction, with the code available on \url{https://github.com/AIPMLab/SICDN}. 
\end{abstract}

\section{Introduction}
\label{1}

In the healthcare field, the quality of medical data varies between hospitals, while medical data sets in different places are characterized by an imbalance in size and quality \cite{griot2024impact}. And the use of deep learning (DL) models to aid in diagnosis, prediction and prognosis in medical tasks has been rapidly developed, but the performance of DL models can be influenced by the size and quality of the data set \cite{10385298}. In addition, medical data processing follows strict regulations, such as the General Data Protection Regulation (GDPR) \cite{albrecht2016gdpr}, making the data used for centralized ML unshareable, and the requirement for large amounts of data cannot be met \cite{hu2023potential}.

To address data acquisition challenges, Federated Learning (FL) \cite{chaddad2023explainable} is a prominent distributed solution, although resource-intensive. Feature selection techniques such as L1 regularization enhance training efficiency, but may not fully model nonlinear relationships \cite{kang2023improved}. Principal Component Analysis (PCA) on CNN features, as explored in \cite{chaddad2023texture}, efficiently summarizes texture data, reduces dimensionality, and mitigates overfitting, despite its computational demands. Attention mechanisms offer dynamic feature selection, yet require careful parameter tuning and extensive training \cite{chen2023combining}.

We introduce a model refinement technique using SHAP (Shapley Additive exPlanations) \cite{lundberg2017unified}, one technique of explainable artificial intelligence (XAI) that quantifies the contribution of each input feature \cite{chaddad2024generalizable}, to boost the interpretability of CNNs for classification tasks by assessing each feature impact. This method is tailored for limited datasets, allowing models to identify unique features efficiently without extensive data. It is especially beneficial in medical diagnostics, adapting to limited data variations for personalized treatment plans. We present the SHAP-integrated convolutional diagnostic network (SICDN), which refines CNNs by integrating SHAP values into the weight update process, focusing on the fully connected layer to simplify memory usage and enhance model efficiency. We also factored in historical model weights and hyperparameters to aid in weight updates. The key contributions of the paper are summarized as follows.

\begin{itemize}
    \item We propose a SICDN algorithm to improve the performance metric for limited datasets such as medical images. We demonstrate the superiority of SICDN through experiments in data sets of pneumonia and breast cancer.
     \item We introduce a new method for updating CNN weights and a new model architecture by expanding SICDN. We also explore how hyperparameter settings affect model performance.
\end{itemize}

\section{Methodology}\label{S3}
The training process of the proposed SICDN is illustrated in Figure \ref{fig:shap}. The training process is as follows:

\begin{figure*}[!ht]    \includegraphics[width=0.98\textwidth]{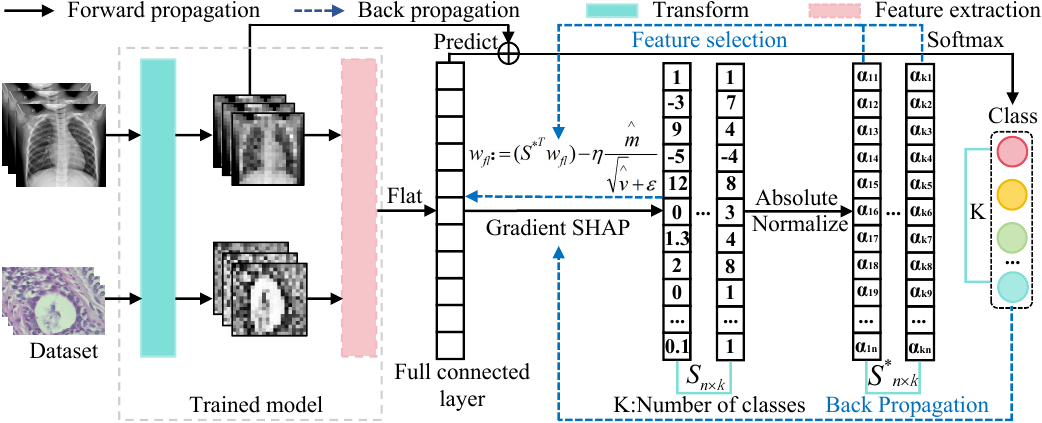}
    \centering
        \caption{Flowchart of the training for SICDN. (1) Transform the medical images and extract features using the trained model, (2) Flatten the features and compute the Shapley values, (3) Normalize the absoluted Shapley value, (4) Use the normalized Shapley values as an importance matrix to multiply with the weights matrix of the fully connected layer and update the weights.}
    \label{fig:shap}
\end{figure*}

\subsection{Feature extraction and selection}

We use densely connected convolutional networks-121 (DenseNet-121) \cite{huang2017densely} to extract features from medical images. And its dense connections and feature reuse lead to efficient training and better image classification, as proven by our comparative experiments. We use the Gradient SHAP interpreter \cite{sundararajan2017axiomatic} 
to calculate the importance of features of the fully connected layer. Specifically, the gradient SHAP computes the Shapley values by integrating model gradients along a path function between background samples \( a' \) and target samples \( a \). The path is defined as \(g(a', a, \alpha) = a' + \alpha \cdot (a - a') + \epsilon \), where \(\alpha\) interpolates between \(a'\) (\(\alpha=0\)) and \(a\) (\(\alpha=1\)), and \(\epsilon\) is random noise. The Shapley values \(\phi_i\) are calculated as Eq. \ref{equation5}:
\begin{equation}
\phi_i = \mathbb{E}_{a' \sim \text{data}} \left[ \int_{0}^1 \frac{\partial f(g(a', a, \alpha))}{\partial a_i} d\alpha \cdot (a_i - a_i') \right]
\label{equation5}
\end{equation}

We denote the set of features of the fully connected input layer after flatting as \( \{x_i\}_{i=1}^n \). The number of input features of the fully connected layer of the DenseNet-121 is 1024. We assume that the number of input features \( x_i\) of the fully connected layer is \( n \), then for a multiclassification task with the number of classes \( k \), we compute Shaply values \( n \times k \), denoted as Shaply value matrix \( S_{n \times k} \), which corresponds to the importance of the features in the different classes. A positive Shaply value means that the feature plays a positive role in the prediction of the class, and a negative value means that it plays a negative role, and the higher the absolute value of the Shaply value, the stronger the effect of the feature on the prediction of the model. For the binary classification task, a negative Shaply value for a feature \( x_i \) of class \( 1 \) means that \( x_i \) acts negatively for class \( 1 \) while acting positively for class \( 2 \). Therefore, for the binary classification task, each element in the Shaply value matrix can be reserved for selecting features that play an important role in model prediction.

We first take the absolute value of the Shaply value matrix \( S_{n \times k} \) in binary classification tasks. The batch size is set for this model. If Shaply values are calculated with \( m \) samples as a batch, the average values are taken as shown in Eq.\ref{average_shap}:

\begin{equation}
S'_{ij}  = |\frac{1}{m} \sum_{k=1}^m S_{ijk}|
\label{average_shap}
\end{equation}

We normalize all elements in the matrix \(S'\) to \((0, 1)\) as expressed in Eq. \ref{np}. The processed matrix is denoted as \(S_{n \times k}^{*}\). If the maximum Shaply value \(max(S')\) and the minimum Shaply value \(min(S')\) of the features are equal, then all elements of the matrix \(S'\) are normalized to \(1\). This means all features contribute to the prediction result equally, and feature selection is unnecessary. In addition, the normalization process can mitigate skewness in feature importance values.

\begin{equation}
S^*_{ij} = 
\begin{cases} 
1 & \text{if} \max(S') = \min(S') \\
\frac{S'_{ij} - \min(S')}{\max(S') - \min(S')} & \text{else}
\end{cases}
\label{np}
\end{equation}

Consequently, the number of weights of the fully connected layer connections is \(n \times k\). We transpose \(\ S_{n \times K}^{*} \) and multiply it by the weights of the fully-connected layer \(W_{fl}\) to achieve the feature selection process as in Eq.\ref{fl}. This concludes the feature selection process, and a new round of weight updating begins next.

\begin{equation}
W_{fl} := (\mathbf{S}^{*T} W_{fl}) 
\label{fl}
\end{equation}

\subsection{Weights update}
The most common weight update method in CNNs is the gradient descent (GD) method \cite{ruder2016overview}. However, we choose the Adaptive Moment Estimation (Adam)  optimizer for weight updating \cite{kingma2014adam}. Adam has been widely used in recent studies for its adaptive learning rate, momentum and bias correction, and fast convergence \cite{kabiri2024amadam}. Compared to GD, Adam combines the advantages of momentum and Root Mean Square Propagation (RMSProp). It computes the first- and second-order momentum of the gradient to dynamically adjust the learning rate. The specific network weight update process of the proposed SICDN is as follows. First, the gradient of the fully connected layer weights \(W_{fl}\) is calculated for each time step \(t \) as shown in Eq. \ref{g_t}, where L is the loss function.

\begin{equation}
g_t = \nabla_{W_{fl}} L(W_{fl}) \\
\label{g_t}
\end{equation}

Subsequently, according to Adam's weight update mechanism, based on the \(g_t \), calculate the first-order momentum \(v_t \) and the second-order momentum \(m_t \). Finally, the momentum is used for network weight updating, as shown in Eq. \ref{wupdate}. The constant \(\epsilon\) is a very small value to prevent the denominator from being zero. \(W_{fl}\) denotes the weight matrix of the fully connected layer. 

\begin{equation}
W_{fl} := (\mathbf{S}^{*T} W_{fl}) - \eta \frac{\hat{m}_t}{\sqrt{\hat{v}_t} + \epsilon}
\label{wupdate}
\end{equation}

As for loss function, we use the cross-entropy loss function as in Eq. \ref{loss}:

\begin{equation}
L(y, p) = - \sum_{i=1}^{M} y_i \cdot log(p_i)
\label{loss}
\end{equation}

\subsection{Historical weighted moving average}

SICDN screens features, reducing available features for the model. To address this, we introduce historical weights using an exponentially weighted moving average for feature selection. We experiment with different $\lambda$ values to analyze the impact of SHAP method shares on model performance, and normalize fully connected layer weights as described in Eq. \ref{normalize_W}:

\begin{equation}
W^*_{fl_{ij}} = 
\begin{cases} 
1 & \text{if } \max(W) = \min(W) \\
\frac{W_{fl_{ij}} - \min(W)}{\max(W) - \min(W)} & \text{else}
\end{cases}
\label{normalize_W}
\end{equation}

Subsequently, SHAP and historical weights are jointly updated with a new round of weights for the fully connected layer \(W_{fl}\), with the same process as in Eq. \ref{Historical}:

\begin{equation}
W_{fl} := ((\lambda \mathbf{S}^{*T} )+(1 - \lambda)W_{fl}^{*T}) W_{fl}- \eta \frac{\hat{m}_t}{\sqrt{\hat{v}_t} + \epsilon}
\label{Historical}
\end{equation}
where $\lambda$ is the hyperparameter representing the proportion of the feature selection process using the SHAP method.

\section{Experiments}\label{S4}
\subsection{Dataset}

\textbf{1) Breast cancer}: This data set was taken from \cite{spanhol2015dataset} and consists of 1693 images of breast tumor pathology. These images are divided into two classes, with 547 benign and 1146 malignant. \textbf{2) Pneumonia }: We used a publicly available Kaggle dataset \cite{Dataset_Chestray} that was taken from a large collection of labeled Optical Coherence Tomography (OCT) and Chest X-Ray images \cite{kermany2018identifying}. The data set included 5856 images, divided into two classes: 4273 pneumonia and 1583 normal samples.

The division of the different datasets with detailed sample sizes is reported in Table \ref{number}. 

\begin{table} [!ht]
    \centering
    \setlength{\tabcolsep}{0.47cm}
    \caption{Number of samples for training, validation, and test sets for the three datasets.}
    \dorowcolors
    \begin{tabular}{ccccc}
    \toprule
         Classes & Train & Validation & Test  & Total \\
           \midrule
         \multicolumn{5}{c}{\textit{Complete pneumonia dataset}} \\
          \midrule
    \rowcolor{white} Pneumonia  &3875& 214& 184 & 4273\\
    \rowcolor{gray!15}Normal  & 1341 & 138 & 104 & 1583 \\
     \midrule
     \rowcolor{white}\multicolumn{5}{c}{\textit{Subset pneumonia}} \\
     \midrule
     \rowcolor{white}Pneumonia & 357 & 42 & 44 & 443\\
     \rowcolor{gray!15}Normal & 340& 40 & 40 & 420\\
    \midrule
    \rowcolor{white}\multicolumn{5}{c}{\textit{Breast cancer}} \\
    \midrule
    \rowcolor{white}Benign & 345 & 26 & 176 & 547\\
    \rowcolor{gray!15}Malignant & 726 &51 & 369& 1146\\
    \bottomrule
    \end{tabular}
    \label{number}
\end{table}

\subsection{Experimental setup}

We designed SICDN for efficient handling of limited datasets, comparing its performance on complete and subset pneumonia datasets with a \(10 : 1\) training set ratio using metrics, AUC, accuracy, recall, and F1 score. We also assessed its generalization across different datasets and optimized model settings by tuning the $\lambda$ value for a weighted exponential moving average of historical network weights. The experiments were conducted on an RTX 4090 GPU and an Intel i9-13900K CPU, using PyTorch for implementation.

We use Densenet-121 as our network backbone \cite{huang2017densely}. We compare SICDN with four state-of-the-art CNN models, namely DenseNet-121, ResNet-50 \cite{he2016deep}, Edge Convolutional Neural Network Next Base (EdgeNeXt-Base) \cite{maaz2022edgenext} and convolutional network next base (ConvNeXt-base) \cite{liu2022convnet}. The comparison between DenseNet-121 and SICDN serves as an ablation study to assess the effectiveness of the SHAP method. Our experiments are divided into three parts:

\textbf{1)} Pneumonia classification using both the full and subset pneumonia datasets to verify SICDN's effectiveness with limited data.

\textbf{2)} Classification tasks on breast cancer and subset pneumonia datasets to evaluate SICDN’s generalizability.

\textbf{3)} Explore the impact of the value of $\lambda$ on the performance of the model in subset pneumonia.

For all experiments, the parameter settings were the same to ensure comparable results, with the batch size set to 8, the epochs set to 100 and Adam optimizer. All parameter choices were based on our pre-experimental validation. SICDN aims to select the  model with  highest accuracy on the validation set as the pre-trained model for feature selection.

\begin{table}[!ht]
    \centering
\setlength{\tabcolsep}{0.25cm}

    \caption{A summary of the top (average) performance metrics between the three datasets in classification tasks.}
    \dorowcolors
    
    \begin{tabular}{cccc}
    \toprule
         Model  & Accuracy & Recall  & F1 \\
           \midrule
         \multicolumn{4}{c}{\textit{Complete pneumonia dataset}} \\
          \midrule
    \rowcolor{white}\ DenseNet-121    & 83.33 (75.97)& 100.00 (99.03) & 87.37 (84.05)\\			 
    \rowcolor{gray!15}ResNet-50        & \textbf{86.81 (76.25)}& 100.00 (98.41) & \textbf{90.21 (84.13)}\\
    \rowcolor{white}\ EdgeNeXt-Base    & 76.39 (65.73)& 100.00 (98.08) & 82.71 (78.48)\\
    \rowcolor{gray!15}ConvNeXt-base     & 82.99 (76.79)& 100.00 (99.08)& 87.96 (84.53) \\

    \rowcolor{white}Ours              & 80.90 (78.47) & 100.00 (99.32)	 & 86.67 (85.50) \\		
        \rowcolor{gray!15}\ $\lambda$=0.5    & 83.68 (76.64)& \textbf{100.00 (99.37)} & 88.62 (88.48)\\			
     \rowcolor{white}$\lambda$=0.6     & 82.64 (75.34) & 100.00 (99.26)& 87.92 (93.75) \\
     \midrule
     \rowcolor{white}\multicolumn{4}{c}{\textit{Subset pneumonia}} \\
     \midrule
     \rowcolor{white}\ DenseNet-121   & 95.24 (85.39)& 100.00 (98.02) & 95.65 (87.42)\\
    \rowcolor{gray!15}ResNet-50        & 90.48 (82.05)& 100.00 (95.84) & 91.49 (84.15) \\
    \rowcolor{white}\ EdgeNeXt-Base    & 80.95 (66.14)& 100.00 (92.95)& 82.83 (73.30)\\
    \rowcolor{gray!15}ConvNeXt-base     & 83.33( 75.33) & 100.00 (96.39) &85.71 (80.75) \\

        \rowcolor{white}Ours       & \textbf{97.62 (88.58)} & 100.00 (98.11)	 & \textbf{97.73 (89.67)}\\		
    \rowcolor{gray!15}\ $\lambda$=0.5    & 96.43 (88.45)& 	\textbf{100.00 (98.59)}  & 	96.7 (90.02)  \\			
     \rowcolor{white}$\lambda$=0.6    & 95.24 (90.43)& 100.00 (97.8)& 	95.65 (91.36) \\				
    \midrule
    \rowcolor{white}\multicolumn{4}{c}{\textit{Breast cancer dataset}} \\
     \midrule
     \rowcolor{white}\ DenseNet-121    & 95.78 (92.10)& 98.37 (95.45) & 96.91 (94.26)\\
    \rowcolor{gray!15}ResNet-50         & 96.33 (92.28)& 98.64 (95.54) & 97.30 (94.40) \\
    \rowcolor{white}\ EdgeNeXt-Base   & 86.61 (84.89)& 97.02 (92.59) & 90.48 (89.20)\\
    \rowcolor{gray!15}ConvNeXt-base    & 91.74 (87.30) & 97.56 (92.69) & 93.88 (90.85) \\

    \rowcolor{white}Ours            &\textbf{97.06 (94.25)}& 99.73 (97.37) & \textbf{97.84 (95.87)} \\			
    \rowcolor{gray!15}\ $\lambda$=0.5&96.88 (92.98)&\textbf{100.00 (96.85)} & 97.71 (95.05)\\			
     \rowcolor{white}$\lambda$=0.6    & 96.88 (93.38) &100.00 (96.70) & 97.71 (95.27) \\			
    \bottomrule
    
    \end{tabular}
     \raggedright
    {Bold values indicate the best model performance in the corresponding dataset}
    \label{tab:result}
\end{table}

\begin{figure}
    \centering
    \includegraphics[width = 0.48 \textwidth]{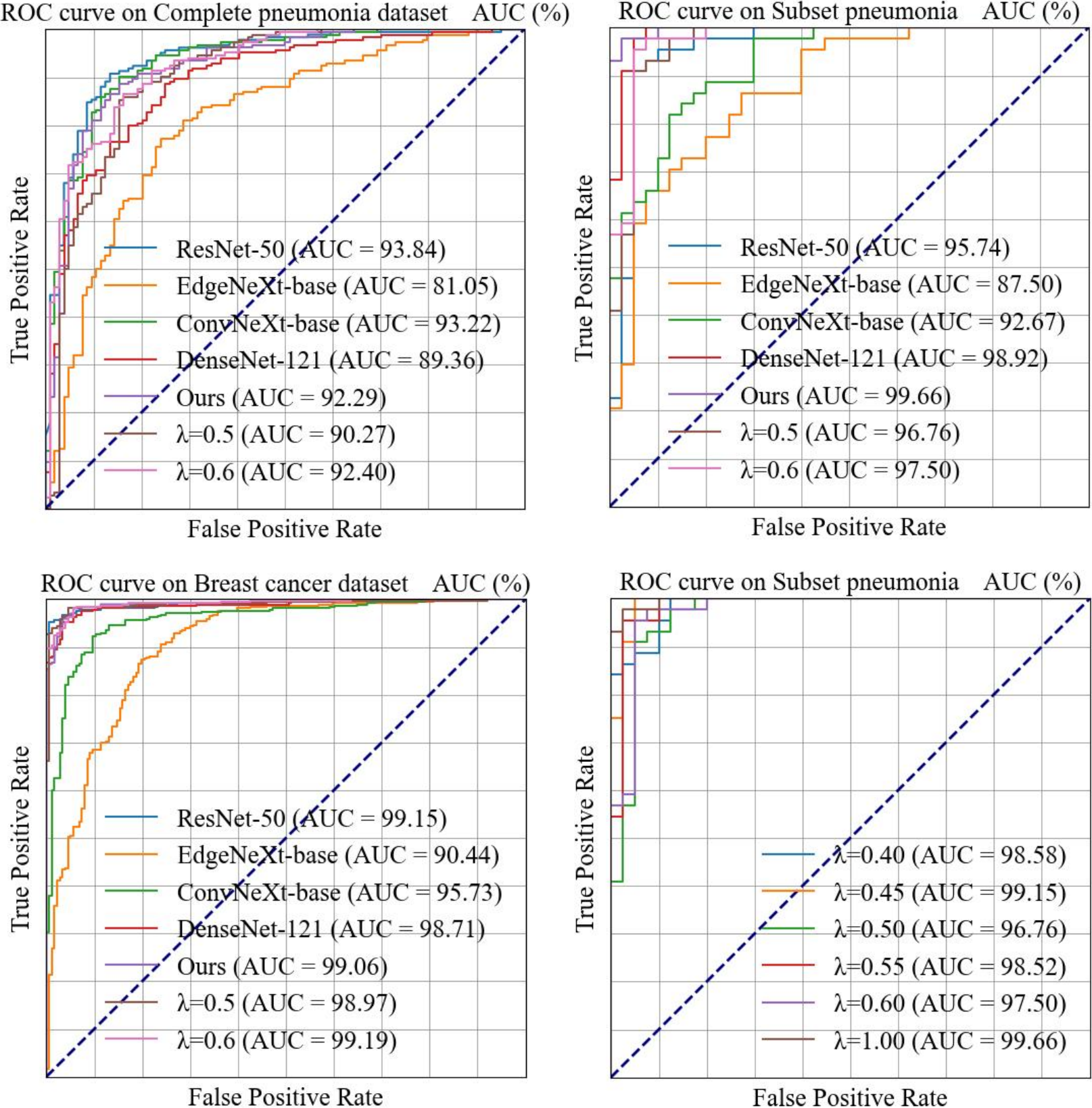}
    \caption{The ROC curves and top AUC scores of the models.}
    \label{fig5}
\end{figure}

\subsection{Results}

The test results of four CNNs, SICDN, and its extension model in the three datasets are reported in Table \ref{tab:result}. The values are presented as top (average). We choose the highest epoch result in the test set as the top value. The average of 100 epoch results is taken as the average value. The top value shows the best performance of the model. The average value indicates the stability of the model during the training process. Figure \ref{fig5} shows the ROC curves and top AUC scores of the models. 

From Table \ref{tab:result} and Figure \ref{fig5}, for the complete pneumonia data set, ResNet-50 achieved the best performance compared to other models, with almost all performance metrics the highest: AUC: 93.84\%, accuracy: 86.81\%, and F1 score: 90.21\%. In the subset pneumonia, SICDN exhibited the best performance, with an AUC score of 99.66\%, accuracy of 97.62\%, and F1 score of 97.73\%. These three evaluation metrics were nearly 17\% higher than those of EdgeNeXt-base and nearly 2\% higher than those of DenseNet-121. SICDN showed optimal performance in almost all metrics. Similarly, in the breast cancer dataset, SICDN achieved the best accuracy and F1 score of 97.06\% and 97.84\%, respectively. The extended model of SICDN (\(\lambda=0.6\)) achieved the best AUC score of 99.19\%. The extended models (\(\lambda=0.5\) and \(\lambda=0.6\)) also performed excellently, especially in terms of recall and F1 scores, demonstrating the robustness of SICDN. Table \ref{tab:pararesult} reports the performance metrics of the SICDN extension models for the \(\lambda\) values. When \(\lambda=1\), the model performs excellently in all metrics, particularly in AUC (99.66\%), accuracy (97. 62\%) and F1 score (97.73\%), achieving the highest values. Models with other \(\lambda\) values show lower performance metrics.
\begin{table} [!ht]
    \centering
    \setlength{\tabcolsep}{0.43cm}
    \caption{A summary of the top (average) performance metrics between a range of $\lambda$ values settings on subset pneumonia.}
    \dorowcolors
    \begin{tabular}{cccc}
    \toprule
         $\lambda$  & Accuracy & Recall  & F1 \\
           \midrule
    \rowcolor{white}0.40     & 95.24 (84.27)& \textbf{100.00 (99.95)} & 	95.65 (87.11)\\		 
    \rowcolor{gray!15}0.45   & 95.24 (86.79)& 100.00 (98.61)	 & 95.65 (88.76)\\
    \rowcolor{white}0.50    & 96.43 (88.45)&100.00 (98.59)	& 96.70 (90.02)\\	
    \rowcolor{gray!15}0.55   &95.24 (86.39)& 100.00 (98.70)	& 95.65 (88.49)\\				
    \rowcolor{white}0.60  & 95.24 (90.43)& 100.00 (97.80)& 95.65 (91.36)\\
    \rowcolor{gray!15}1.00     & \textbf{97.62 (88.58)} & 100.00 (98.11) & \textbf{97.73 (89.67)}\\		
    \bottomrule
    \end{tabular}
 \raggedright
{Bolded values indicate the best model performance}
    \label{tab:pararesult}
\end{table}

Overall, SICDN performs better on limited datasets (subset pneumonia and breast cancer dataset) with high robustness. In a large data set (complete pneumonia data set), model evaluation metrics are still competitive. One of the advantage of this model that can be used with FL model \cite{10431827}, XAI \cite{chaddad2024generalizable} and radiomics \cite{10433679} for classification tasks.

\section{Conclusion}\label{S5}

This paper presented SICDN to address data access challenges from privacy regulations, showing its superiority on limited datasets compared to four popular CNNs based on experimental results. Future research could integrate SICDN with FL to securely gather diverse data and personalize the global model for local data distributions. Additionally, extending SICDN to other CNNs represents an important future direction to further validate the generalizability of SICDN.

\section*{Acknowledgment}
This research was funded by the National Natural Science Foundation of China (\#82260360), Innovation Project of GUET Graduate Education (\#2024YCXS196), the Guilin Innovation Platform and Talent Program (\#20222C264164), and the Guangxi Science and Technology Base and Talent Project (\#2022AC18004, \#2022AC21040).


\bibliographystyle{IEEEbib}
\bibliography{refs}

\end{document}